# Canonical distribution and incompleteness of quantum mechanics

V. A. Skrebnev

Kazan Federal University, Institute of Physics, Russian Federation

e-mail: vskrebnev@mail.ru

The paper discusses the physical groundlessness of the models used for the derivation of canonical distribution and provides the experimental data demonstrating the incompleteness of quantum mechanics. The possibility of using statistical ensembles is presented as a consequence of the existence of probabilistic processes which are not accounted for by quantum mechanics. The paper provides a new analytical derivation of canonical distribution for macrosystems which takes into account subquantum processes. The paper discusses the possibility of the experimental study of a probability which is beyond quantum mechanics.

## 1. Critique of the models used for the derivation of canonical distribution

For the derivation of the canonical distribution, statistical mechanics uses models which cannot be physically substantiated within the limits of existing theories and common sense. We will briefly consider these models.

### Statistical matrix and canonical distribution

A well-known textbook [1] allocates a subsystem from an isolated macroscopical system; this subsystem is small in comparison to the entire system, but is also macroscopical. When the subsystem is big enough, the energy of its interaction with the surrounding parts of the system will be small in comparison with its internal energy and it is possible to consider that the subsystem is quasi-isolated and various subsystems of the system are statistically independent.

Let $\psi_m$ be the eigenfunctions of the subsystem when the interaction of the subsystem with its environment is neglected. The subsystem's wave function can be expanded into a sum of eigenfunctions $\psi_m$:

$$\psi = \sum_m c_m \psi_m, \qquad (1)$$

The average value of any quantity $f$ is found by means of the formula

$$\overline{f} = \sum_{n,m} c_n^* c_m f_{nm} \qquad (2)$$

where

$$f_{nm} = \int \psi_n^* \hat{f} \psi_m dq.$$

Chapter 1 of the textbook [1] says that the transition from the quantum-mechanical description of a macrosystem to the statistical one can be viewed, in a sense, as the averaging by various $\psi$-states. Then, the average value of $f$ will be expressed by the formula:

$$\overline{f} = \sum_{m,n} w_{mn} f_{nm} \qquad (3)$$

The set of all values $w_{mn}$ constitutes a density matrix or a statistical matrix.

The solution of the Schrödinger equation

$$i\hbar \frac{\partial \psi}{\partial t} = \hat{H}\psi \qquad (4)$$

for the isolated system can be written as:

$$\psi = \sum_m c_m(t) \psi_m, \qquad (5)$$

where

$$c_m(t) = c_m(0) \exp\left(-\frac{i}{\hbar} E_m t\right).$$

$|c_m(t)|^2$ determines the probability of the system to be in a state with energy $E_m$. Because $|c_m(t)|^2 = |c_m(0)|^2$, we see that quantum mechanics does not allow the system to pass into a state with a set of $|c_m(t)|^2$ different from the initial.

For statistical operator $\hat{w}$ corresponding to the density matrix $w_{mn}$, the use of expression (5) leads to the equation [1]:

$$\frac{\partial \hat{w}}{\partial t} = \frac{i}{\hbar}(\hat{w}\hat{H} - \hat{H}\hat{w}) \qquad (6)$$

In equilibrium, the statistical operator commutates with the Hamiltonian and is diagonal in the energy representation.

The problem of finding the statistical distribution of the quasi-isolated subsystem is reduced in [1] to determining the probabilities $w_n = w_{nn}$, which are the distribution function in the quantum statistics.

Commutating the operator of any quantity with a Hamiltonian means that this quantity is an integral of movement. The distribution function of the system consisting of several quasi-isolated and statistically independent subsystems is a product of distribution functions of these subsystems. Accordingly, the logarithm of the distribution function should be an additive integral of movement. Thus, the logarithm of the distribution function $w_n$ can be written as [1]:

$$\ln w_n = \alpha - \beta E_n, \tag{7}$$

whence we have:

$$w_n = e^{\alpha - \beta E_n} = Z^{-1} e^{-\beta E_n}, \tag{8}$$

$$Z = \sum_n e^{-\beta E_n}$$

The formula (8) represents the canonical distribution.

Chapter 1 of the textbook [1] emphasizes that «the averaging by various $\psi$-states is rather conditional. In particular, it would be absolutely wrong to consider that the system can be with various probabilities in various $\psi$-states, and that the averaging being conducted is the averaging by these probabilities; such statement in general would contradict the main principles of quantum mechanics». No definition of conditional averaging is given, and the need for statistical averaging is explained by the incompleteness of our knowledge about the system. It is obvious that the incompleteness of our knowledge can't explain either necessity, or possibility of the averaging offered in [1].

*Canonical distribution as a consequence of the interaction of the system with its environment*

Since it is impossible to explain the averaging by various $\psi$ –states, books on statistical mechanics use alternative models for deriving canonical distribution (see, for example, [2]). These models consider only the eigenstates of the system to be system states. In one of the models the system is considered to be a part of a Universe in an equilibrium all of whose states are equiprobable (the word "Universe" is usually modestly spelled with quotation marks). Chapter 3 of the textbook [1] also uses this model, and does not refer back to chapter1 where the canonical distribution was derived on the basis of the "conditional" averaging. Another model assumes that the "Universe" consists of a very great number of systems identical to the system under consideration. In both models the system's interaction with its environment is considered extremely weak. At the same time, the transition of the system from one eigenstate to another is considered to be caused by its interaction with the environment. It is obvious that the formula (8) can be used to calculate the observed quantities only if during the measurement the system has time to visit all states of the spectrum repeatedly. However, the aforementioned models, when they arrive at formula (8), do not correlate the values of the contact with the environment, spectral diapason of the system energy and measurement time.

As neither the Universe, nor "Universe" are in equilibrium, nor do they consist of a great number of systems identical to the system under consideration, it is possible to say with good reason that both models have no relation to physical reality. As it is impossible to prove these models from the point of view of physics and common sense, they become not so much an object of science but an object of faith. To doubt faith is intolerable; therefore to change the minds of those who believe is extremely difficult. In the meantime this belief hinders our understanding and description of the phenomena in the microworld and their connection with the natural processes occurring round us.

The speed of arriving at the canonical distribution does not depend on the properties of the surface of the macrosystem, nor on the structure of its environment. Thus, the influence of environment does not explain the transition of the probabilities of the macrosystem's eigenstates to canonical distribution. This means that there must be internal processes which determine the transition of the initial distribution of probabilities to the canonical distribution. Hence, canonical distribution may be derived as a result of internal processes within the macrosystem – the processes not described by the existing quantum formalism.

## *2. Experimental evidence of the incompleteness of quantum mechanics*

The models used for the derivation of canonical distribution do not call into question the absolute accuracy of quantum mechanics. However, one must remember that both classical and quantum mechanics have resulted from the observation of systems with small number of objects. If the number of objects (e.g. particles) in a system is small and calculations are possible, the mechanics show amazing accuracy. One might assume that in systems with macroscopically great number of particles quantum mechanics would also be absolutely exact. However, this assumption contradicts the irreversibility of evolution of the macrosystems, the second law of thermodynamics and the experimental data received on concrete physical objects. Some data are presented below.

### *Time reversal experiments*

Experiments [3, 4] were carried out in the Kazan Federal University, where the sign of the Hamiltonian of an isolated spin system reversed with predetermined accuracy. As the general solution of the Schrödinger equation has the following form:

$$\psi(t) = e^{-i\frac{\hat{H}}{\hbar}t}\psi(0), \qquad (9)$$

it is evident that Hamiltonian sign reversal is identical to time sign reversal and it must result in the system's reversion into initial state. However, it turned out that the experiments [3, 4] cannot be correctly described on the basis of the reversible equations of quantum mechanics. Note that otherwise it would break the second law of thermodynamics.

### *Spin-lattice relaxation in paramagnetics*

To study the transmission of energy of the paramagnetic centers to the oscillations of a crystal lattice, i.e. spin-lattice relaxation (SLR), a saturating impulse leading to the alignment of spin energy level populations is imposed on the system. After the saturating impulse is removed, the equilibrium is established in the spin-phonon system. SLR can be determined by raman processes when the energy of spins is transferred to the entire phonon spectrum of the crystal. In the diluted paramagnetics the thermal capacity of a phonon system is much more than the thermal capacity of spins. Therefore, after the equilibrium is established, the temperature of the spin-phonon system practically doesn't differ from the temperature before the saturating impulse.

As the temperature drops, SLR begins to be determined by one-phonon or direct processes where spin energy is transferred to phonons with the frequencies lying within the resonant curve. If the thermal capacity of resonant phonons is comparable to the spin system thermal capacity, the phonons may heat, leading to the "phonon bottleneck". In this case the change of the spin level population after the removal of the saturating impulse occurs in two stages. At the first stage there is SLR, i.e. a transmission of spin energy to phonons. At the second stage the energy of spin-phonon system is transferred to the thermostat.

The degree of contact of spin-phonon system with the thermostat is essential to the second stage. As to the SLR as such, it doesn't depend on the degree of contact with the thermostat in any way. SLR is well described by the formulas which do not contain the parameters determining the communication of the system with the thermostat (see, for example, formulas (10.45), (10.49), (10.55), (10.58), (10.76) of the book [5]). Hundreds of experiments studying SLR have been conducted in the world. Experiments at low temperatures have placed the sample both in liquid and in gaseous helium of various densities, i.e. at vastly different degrees of influence of the environment on the system. At low density of gaseous helium, it takes hours to establish the equilibrium between the spin-phonon system and the thermostat, while within the spin-phonon system, due to the raman processes, the SLR takes milliseconds and microseconds [5, 6]. No experiment has observed that SLR might be influenced by helium density, i.e. by the degree of interaction with the thermostat, neither for raman, nor for direct processes.

Let's accept as the spin-phonon system's initial state its state after the removal of the saturating impulse. Wave function of this state can be written as:

$$\varphi = \sum_m a_m(0)\, \varphi_m, \qquad (10)$$

where $\varphi_m$ are the eigenfunctions of the spin-phonon system, and zero magnetization corresponds to a set of coefficients $a_m(0)$. The state of the isolated spin-phonon system at the moment *t* is described by function

$$\varphi = \sum_m a_m(t)\, \varphi_m, \qquad (11)$$

where

$$a_m(t) = a_m(0)\, exp\left(-\frac{i}{\hbar} E_m t\right). \qquad (12)$$

The probability of the system state with energy $E_m$ is equal to

$$|a_m(t)|^2 = |a_m(0)|^2 \qquad (13)$$

and it does not change with time.

As a result of SLR, canonical distribution is established in spin-phonon system and the spin energy level populations are determined by the temperature of this system. According to the models of the derivation of the canonical distribution which consider only eigenstates of systems, the establishment of the canonical distribution should mean a transition of the spin-phonon system from the initial state (10) to one of eigenstates with the further transitions between these states under the influence of the thermostat. According to the model offered in chapter 1 of the textbook [1], the initial state should turn into some set of eigenstates whose probabilities are determined by the canonical distribution (8). However, quantum mechanics doesn't allow such transformations for an isolated system. Hence, the transition of the system from the state (10) to canonical distribution would have to be explained only by the influence of the system's environment and by nothing else. Accordingly, the speed of such transition should depend on the degree of this influence. However, as shown above, such dependence is absent. The conclusion follows that the establishment of canonical distribution in the spin-phonon system after removal of the saturating impulse occurs not under the influence of the environment, but due to the internal processes which can't be described by quantum mechanics.

The experimental data presented above show that quantum-mechanical probability does not cover the entire probabilistic nature of the microworld and that God plays dice not exactly the way prescribed by Schrödinger.

Thus, the assumption of absolute accuracy of quantum mechanics for the description of processes in macrosystems appears erroneous. This circumstance allows us to understand and interpret more accurately the physical phenomena in the world surrounding us.

### *3. Canonical distribution as a result of the transitions not following from quantum mechanics*

In compliance with (5) the total system energy may be written as follows:
$$E = \sum_m |c_m(t)|^2 E_m. \tag{14}$$

Normalization requirement gives us the following:
$$\sum_m |c_m(t)|^2 = 1. \tag{15}$$

If the number of levels of the system energy is more than two, equations (14) and (15) have a great number of solutions for $|c_m(t)|^2$. This means that same value of the total energy $E$ corresponds to a great number of different system states which may not be received one from another as a result of temporal system evolution.

In a classic case, a manifold within phase space corresponds to a set of system states with a definite value of energy. We can introduce the space of the coefficients $c_m$ of the system state function expansion. A manifold will correspond to the value of the system's total energy $E$ in this space, as well as in the phase space of a classical system. Let's consider a generalized quantum microcanonical ensemble for the macrosystem as a set of its states in the given manifold, with equal probability of these states.

It is possible to suppose that the average value of $|c_m|^2$ by generalized quantum microcanonical ensemble corresponds to the canonical distribution. It is demonstrated in [7] that in a system with as little as three levels the formula (8) nicely describes the result of averaging of the coefficients $|c_m|^2$ by the generalized quantum microcanonical ensemble. However, analytical and numeric studies in [8] have shown that in systems with a large number of levels of energy the average value of $|c_m|^2$ by generalized quantum microcanonical ensemble is in a significant departure from the Boltzmann-Gibbs statistics. Notably, [8] does not discuss a physical substantiation for the possibility of averaging on the generalized quantum microcanonical ensemble.

Clearly it is only possible to treat $|c_n|^2$ as a probability of the system having energy $E_n$ if the system transitions between states with different $E_n$. It is possible to say that virtual states with energy $E_n$ appear in the system. We will term them $E_{nv}$-states. Then $|c_n|^2$ is proportionate to the average time of the system being in the $E_{nv}$ - state.

Hidden processes in physical systems may be quite fast. Recently [9, 10] have shown that the lower boundary of the speed of the Einstein's "spooky action at a distance" is 10 000 light speeds. Thus it is natural to consider that the frequency of the quantum system's transitions between its $E_{nv}$ - states may be extremely high. Consequently, the time $t_v$ of one visit of the system of any $E_{nv}$ - state will be many orders of magnitude shorter than the times used in the experiment. This will provide for the needed closeness of theoretical and experimental values of $|c_n|^2$.

Experience shows that in a macrosystem left to its own devices after an impact inducing certain initial conditions, the probability of having energy $E_n$ after some time (the relaxation time) becomes described by canonical distribution

$$\rho_n = \frac{e^{-\beta E_n}}{\sum_n e^{-\beta E_n}} \tag{16}$$

The formal derivation of canonical distribution using the method of the most probable distribution assumes that the "Universe" consists of a very great number of systems identical to the system under consideration and distributed along its possible energy levels [2]. This assumption obviously contradicts physical reality. We will use the method of the most probable distribution but instead of a great number of identical systems we will consider a great number of cumulative visits of the system of its $E_{nv}$ - states .

Let $N$ be the number of times the system visits its $E_{nv}$ - states over time $t$, and let $v_n$ be the number of visits of any $E_{nv}$ - state over this time. Obviously,

$$N = \sum_n v_n \qquad (17)$$

Let's introduce the value

$$E_t = \sum_n v_n E_n \qquad (18)$$

Numerous "configurations" determined by various sets of numbers of visits $v_n$ correspond to the value $E_t$. Each configuration may be realized in $P$ ways corresponding to the number of permutations of the visits:

$$P = \frac{N!}{v_1! v_2! \dots v_l! \dots} \qquad (19)$$

Using Lagrange method, we find the maximum of the function $P$ under conditions (17) and (18) and arrive at the most probable value of the numbers of visits of the $E_{nv}$ – state.

$$v_n = \frac{N e^{-\beta E_n}}{\sum_n e^{-\beta E_n}} \qquad (20)$$

The probability $\rho_n$ of the system being in the $E_{nv}$ - state equals the ratio of the number of the visits of this state to the total number of visits $N$; this probability is given by the formula (16). Thus we have arrived at the canonical distribution. The value $\beta$ is determined by the equation

$$E = \sum_n \rho_n E_n \qquad (21)$$

where $E$ is the total energy of the system.

Using (20) it is easy to show that for maximum $P$

$$\ln P_{max} = -N \sum_n \rho_n \ln \rho_n = N(\beta E + \ln \sum_m e^{-\beta E_m}) = \frac{N}{kT}(E - A) \qquad (22)$$

where $A$ is the free energy of the system.

Whereas

$$E - A = TS$$

the entropy of the system is equal to

$$S = k \frac{\ln P_{max}}{N} \qquad (23)$$

It is obvious that the transition of a macrosystem from the initial state to the state with the most probable distribution of the values $v_n$ is only possible if there exist transitions between different system states while the total energy of the system is preserved. Let's call those transitions "relaxation transitions" (r-transitions). It is shown in Section 1 of this paper that canonical distribution cannot be adequately derived as a result of the macrosystem's interactions with its environment. Considering all of the above we view the r-transitions as manifestations of the internal properties of the system, which are not reflected in the existing quantum mechanics formalism. One such property is the tendency of the macrosystem towards the maximum freedom in realizing its state with a given total energy, i.e. towards maximum entropy. Let us note that Botzmann's genius hypothesis regarding the most probable distribution of gas molecules on energy levels has no justification within mechanics. This fact makes Botzmann's hypothesis all the more meaningful.

## *Conclusion*

Undeniably, the equilibrium in a macrosystem is established due to the interaction. It has been proposed in [4] that there exist transitions between different system states while the total energy of the system is preserved; these transitions do not follow from the Schrödinger equation, are not manifested in systems with small number of particles and are the reason for the irreversibility in macrosystem evolution.

The interaction-caused energy exchange between the objects constituting the system within the state described by the $\psi$-function (5) can't lead to a canonical distribution. It follows that the interaction in the systems described by canonical distributions should also cause transitions between states with equal energy, not following from the quantum mechanics. The study of such transitions will further our knowledge of the laws of the microworld.

In macrosystems, the canonical distribution nicely describes the results of experiments even with small measurement time. With the decrease of the number of particles in a system, the probability of r-transitions decreases. This can lead to the discrepancy between the measured values of the system parameters and their values calculated using the canonical distribution. This fact will allow to estimate the value of the probability of r-transitions.

After we face the incompleteness of quantum mechanics, a natural need arises to find the necessary changes and additions to the quantum-mechanical theory. We are confident that the irreversible processes occurring in the macrocosm are connected with r-transitions. Theoretical and experimental study of this connection can become an extremely interesting and promising direction in the development of the science of physics.

One of the examples of the practical value of the ideas and results stated here, is that the existence of the probabilistic processes which do not follow from the Schrödinger equation can lead to serious problems for the attempts to create quantum computers.